\documentclass[conference]{IEEEtran}
\IEEEoverridecommandlockouts
\usepackage{cite}
\usepackage{amsmath,amssymb,amsfonts}
\usepackage{algorithmic}
\usepackage{graphicx}
\usepackage{textcomp}
\usepackage{xcolor}
\usepackage{url}
\usepackage{booktabs}
\usepackage{tabularx}
\def\BibTeX{{\rm B\kern-.05em{\sc i\kern-.025em b}\kern-.08em
    T\kern-.1667em\lower.7ex\hbox{E}\kern-.125emX}}
\begin{document}

\title{Frontend-Assisted Large Audio-Language Models for SLT2026 SmartGlasses Challenge}

%
% Single address.
% ---------------
\author{
\IEEEauthorblockN{Yushi Wang$^{1,2}$, Zeyan Song$^{1,2}$, Xiaobin Rong$^{1,2}$, Chengwei Liu$^{3}$, Jun Gao$^{1,2}$, Jing Lu$^{1,2}$}
\IEEEauthorblockA{$^1$Key Laboratory of Modern Acoustics, Nanjing University, Nanjing, China \\
$^2$NJU-Horizon Intelligent Audio Lab, Horizon Robotics, Beijing, China \\
$^3$Qwen Business Unit of Alibaba, Hangzhou, China \\
\small\texttt{\{yushi.wang, zeyan.song, xiaobin.rong\}@smail.nju.edu.cn}\\
\small\texttt{
liuchengwei.lcw@alibaba-inc.com, 
jun.gao@smail.nju.edu.cn, 
lujing@nju.edu.cn}
}
}
% \author{\IEEEauthorblockN{1\textsuperscript{st} Yushi Wang}
% \IEEEauthorblockA{\textit{Key Laboratory of Modern Acoustics} \\
% \textit{Nanjing University}\\
% Nanjing, China \\
% yushi.wang@smail.nju.edu.cn}
% \and
% \IEEEauthorblockN{2\textsuperscript{nd} Zeyan Song}
% \IEEEauthorblockA{\textit{Key Laboratory of Modern Acoustics} \\
% \textit{Nanjing University}\\
% Nanjing, China \\
% zeyan.song@smail.nju.edu.cn}
% \and
% \IEEEauthorblockN{3\textsuperscript{rd} Xiaobin Rong}
% \IEEEauthorblockA{\textit{Key Laboratory of Modern Acoustics} \\
% \textit{Nanjing University}\\
% Nanjing, China \\
% xiaobin.rong@smail.nju.edu.cn}
% \and
% \IEEEauthorblockN{4\textsuperscript{th} Chengwei Liu}
% \IEEEauthorblockA{\textit{Qwen Business Unit of Alibaba} \\
% \textit{Alibaba Group}\\
% Hangzhou, China \\
% liuchengwei.lcw@alibaba-inc.com}
% \and
% \IEEEauthorblockN{5\textsuperscript{th} Jun Gao}
% \IEEEauthorblockA{\textit{Key Laboratory of Modern Acoustics} \\
% \textit{Nanjing University}\\
% Nanjing, China \\
% jun.gao@smail.nju.edu.cn}
% \and
% \IEEEauthorblockN{6\textsuperscript{th} Jing Lu}
% \IEEEauthorblockA{\textit{Key Laboratory of Modern Acoustics} \\
% \textit{Nanjing University}\\
% Nanjing, China \\
% lujing@nju.edu.cn}
% }

\maketitle

\begin{abstract}
This paper describes our systems for the Track 1 time-stamped speaker-attributed ASR task of the SmartGlasses Challenge. To handle overlapping speech, background noise and far-field conditions in two-person conversations, we employ SoulX-Transcriber, a multimodal large language model that takes full-length audio as input and generates structured transcriptions. In addition, we investigate the effect of speech frontend processing, including a rule-based Guided Source Separation method and a data-driven speech separation model, both of which improve performance on the validation set. Considering inference stability, our final submission uses the fine-tuned SoulX-Transcriber system, achieving a tcpCER of 6.22 on the test set.
\end{abstract}


\section{Introduction}
The SmartGlasses Challenge\footnote{\url{https://aslp-lab.github.io/SmartGlasses/}} aims to promote research on multimodal-LLM-based Time-Stamped Speaker-Attributed ASR (TSA-ASR) and Spoken Language Understanding (SLU). The TSA-ASR task of Track 1 in the SmartGlasses Challenge focuses on generating time-stamped, speaker-attributed transcriptions from four-channel audio captured by smart glasses for two-person everyday conversations with overlapping speech, background interference, and frequent speaker turns.

Recent studies on multimodal large language models have attempted to unify ASR, speaker diarization, and timestamp prediction within a single structured generation framework, including MOSS Transcribe Diarize\cite{MOSS}, VibeVoice-ASR\cite{vibevoice}, SoulX-Transcriber\cite{soulx}, and diarization-aware LLM-based multi-speaker ASR systems\cite{DMASR}. These studies demonstrate the strong potential of end-to-end structured generation. However, most of these systems do not explicitly model or separate overlapped speech, but instead rely on the backend large model to directly transcribe overlapping regions. This paradigm may be insufficient for minimizing word error rate (WER) or character error rate (CER). Moreover, many existing systems are designed for single-channel input, while wearable devices such as smart glasses are typically equipped with multiple microphones that provide useful spatial information.

Motivated by these observations, this report describes our systems for the Track 1 TSA-ASR task of the SmartGlasses Challenge. Our goal is twofold: first, to perform supervised fine-tuning on an open-source large audio-language model so that it better adapts to real recordings in the challenge scenarios; and second, to investigate multi-channel frontend processing methods, including rule-based beamforming and data-driven neural network approaches, for target-speaker extraction or speech separation. For the final submission, we use the most stable system, the fine-tuned version of SoulX-Transcriber, which achieves a tcpCER of 6.22 on the test set.

\section{System Description}
\begin{figure}[h]
\centering
\includegraphics[width=\linewidth]{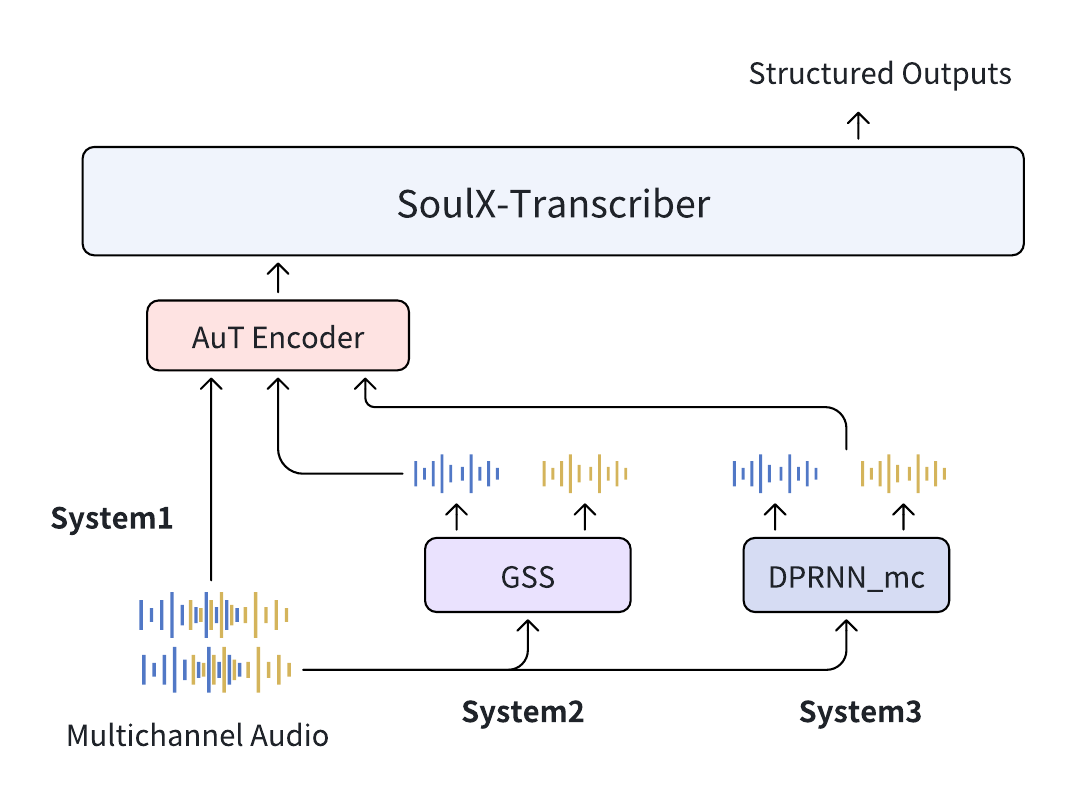}
\caption{Overview of our systems, where the three systems are developed and evaluated independently.}
\label{overview}
\end{figure}

As shown in Fig.~\ref{overview}, based on the open-source SoulX-Transcriber model\footnote{\url{https://github.com/Soul-AILab/SoulX-Transcriber}}, we build three systems. System 1 directly applies SoulX-Transcriber to the channel-averaged mixture speech. Before the LLM backend, System~2 employs Guided Source Separation (GSS)\cite{GSS_1}\cite{GSS} for target-speaker extraction while System~3 uses an improved DPRNN-based\cite{DPRNN} multi-channel separation network to produce two speaker-specific streams. These three systems are developed and evaluated independently.

\subsection{SoulX-Transcriber Backbone}
SoulX-Transcriber is built upon the Qwen3-Omni\cite{qwen3} framework and inherits its long-context audio understanding and autoregressive text generation capabilities. It can generate time-stamped speaker-attributed transcriptions.

SoulX-Transcriber is trained in two stages. First, speaker-aware multi-task continuous pre-training is performed on large-scale pseudo-labeled real conversations and simulated multi-speaker dialogue data. This stage jointly optimizes multiple speaker-related tasks, such as speaker turn prediction, speaker diarization, speaker verification, and ASR, to enhance multi-speaker conversational understanding. Second, the model is supervised fine-tuned on approximately 1,000 hours of high-quality annotated Speaker Diarization and Recognition (SDR) data, improving instruction following, timestamp prediction, speaker attribution, and transcription robustness.

In System 1, SoulX-Transcriber-pretrained denotes the original open-source model, while SoulX-Transcriber-finetuned refers to the model supervised fine-tuned from the open-source checkpoint using the SmartGlasses Challenge training set. 

\subsection{Guided Source Separation Module}
To improve the robustness of transcription in overlapped and far-field speech regions, we introduce Guided Source Separation (GSS)\cite{GSS_1}\cite{GSS} as a multi-channel speech enhancement frontend before the SoulX-Transcriber backend, which forms our System 2.

GSS is a target-speaker extraction method that uses pre-computed speaker activity information to guide blind source separation and beamforming. For each full-length recording, we use speaker-attributed VAD labels to construct target-speaker segments with surrounding context. Following the GSS procedure, the multi-channel waveform is transformed into the STFT domain, followed by WPE dereverberation, CACGMM-based mask estimation, and MVDR beamforming to extract the target-speaker signal. The smart-glasses wearer and the conversation partner are selected as the target speaker in turn, resulting in two single-speaker audio tracks with zero-padded inactive regions. These tracks are independently decoded by SoulX-Transcriber, and the two transcriptions are finally merged into the complete dialogue output.

\subsection{Multi-channel Speech Separation Module}
System~3 introduces an improved multi-channel speech separation model based on the DPRNN backbone, denoted as DPRNN\_mc. The model takes the long-form mixed multi-channel recording as input and outputs two separated audio streams, corresponding to the smart-glasses wearer and the conversation partner, respectively.

Given a four-channel microphone recording, each channel is transformed into the STFT domain, and an Equivalent Rectangular Bandwidth (ERB) filter bank is applied to reduce spectral redundancy while preserving low-frequency resolution. The ERB features from different channels are encoded independently and then fused into a unified multi-channel representation. The fused representation is processed by a stack of Dual-Path Recurrent Neural Network (DPRNN) blocks, which model both local and long-range temporal-frequency dependencies. The decoder predicts two complex-valued time-frequency masks for the two speakers. After inverse ERB transformation, the masks are applied to the reference microphone spectrum, and inverse STFT is used to reconstruct the separated single-channel signals. The two output streams are then independently fed into SoulX-Transcriber.

We train the network with a multi-term time-frequency domain loss. 
For each output stream, given the estimated complex spectrogram $\hat{S}$ and the target complex spectrogram $S$, we define a complex spectral loss as the mean squared error (MSE) between the real and imaginary parts:
\begin{equation}
\mathcal{L}_{\mathrm{comp}}
=
\mathrm{MSE}\left(\Re(\hat{S}), \Re(S)\right)
+
\mathrm{MSE}\left(\Im(\hat{S}), \Im(S)\right).
\end{equation}
In addition, we define a magnitude spectral loss as the MSE between the corresponding magnitude spectrograms:
\begin{equation}
\mathcal{L}_{\mathrm{mag}}
=
\mathrm{MSE}\left(|\hat{S}|, |S|\right).
\end{equation}

To further improve perceptual quality, we incorporate the PMSQE (Perceptual Metric for Speech Quality Evaluation) loss $\mathcal{L}_{\mathrm{pmsqe}}$, which is computed on the time-domain waveforms. The overall loss for a single output is a weighted sum of these three loss functions. Specifically, $\mathcal{L}_1$ denotes the loss for the wearer stream, and $\mathcal{L}_2$ denotes the loss for the partner stream. 
The final training objective is defined as:
\begin{equation}
\mathcal{L}_{\mathrm{total}}
=
\mathcal{L}_1
+
2 \cdot \mathcal{L}_2,
\end{equation}
which assigns a larger weight to the conversation partner stream. To prevent leakage from the interfering speaker, we use VAD labels to replace the waveform regions where the target speaker is inactive in each stream with zeros. Finally, SoulX-Transcriber independently transcribes the two streams and merges their contents.

\subsection{Speaker-Attributed VAD Estimation}
We used the multi-channel DiariZen-based\cite{diarizen} diarization model with ChannelAttention (ChAtt)~\cite{Diarizen_mc} to obtain speaker-attributed VAD labels. This model extends a WavLM-based EEND-VC diarization pipeline by inserting cross-channel communication modules into the early WavLM layers, where ChannelAttention performs multi-head self-attention across microphone channels to capture spatial cues. The resulting local speaker activity predictions are further processed by speaker embedding clustering to produce global speaker-attributed VAD labels for downstream ASR.

\section{Experiments}

\subsection{Datasets and Experimental Setup}
SoulX-Transcriber has a 30B-parameter MoE Thinker and a 0.6B-parameter AuT encoder, while DPRNN\_mc has 4.7M parameters. For supervised fine-tuning of SoulX-Transcriber, we only used the official challenge dataset, which contains approximately 33 hours of speech. A DiariZen model\footnote{\url{https://huggingface.co/BUT-FIT/diarizen-wavlm-large-s80-md-v2}} is extended to multi-channel input using the ChAtt method and trained on the same data.
For training the multi-channel speech separation model, we used real RIRs from the CHiME-8 MMCSG Challenge recorded with Aria glasses and simulated RIRs from the MCAS dataset\cite{MMCSG}, selecting the four glasses-mounted microphone channels from both datasets. We also generated additional four-channel simulated RIRs based on the SmartGlasses microphone layout. The speech and noise data used for simulation were obtained from the URGENT Challenge\cite{URGENT}.

For supervised fine-tuning of SoulX-Transcriber, we adopted LoRA (Low-Rank Adaptation) to perform parameter-efficient adaptation of the language model backbone. The LoRA rank was set to 8, with a scaling factor of $\alpha=32$. For Systems~2 and~3, we first generated enhanced or separated speech from the SmartGlasses training set using GSS and DPRNN\_mc, respectively, and then used the processed speech to fine-tune SoulX-Transcriber. On the validation set, we manually set the decoding configuration of SoulX-Transcriber to temperature = 0.2, top-p = 0.9, top-k = -1, and repetition penalty = 1.05. All experiments were conducted on four NVIDIA 4090 GPUs with 48 GB memory.

\subsection{Experimental Results}

\begin{table}[t]
\vspace{-5pt}
\centering
\caption{Experimental results on the Track 1 TSA-ASR task official validation set.}
\label{tab:exp}
\small
\begin{tabular}{@{}lccc@{}}
\toprule
System & tcpCER & cpCER & DER \\ 
\midrule
Baseline VibeVoice-ASR & 14.09 & 12.88 & 17.24 \\
System 1 SoulX-pretrained & 9.02 & 8.45 & 3.70 \\
System 1 SoulX-finetuned & 7.82 & 7.30 & 2.56 \\
System 2 GSS-SoulX &  &  &  \\
\qquad w/ oracle VAD & 7.39 & 6.65 & 1.14 \\
\qquad w/ estimated VAD & 6.97 & 6.12 & 2.62 \\
System 3 DPRNN\_mc-SoulX &  &  &  \\
\qquad w/ oracle VAD & 7.04 & 6.83 & 0.15 \\
\qquad w/ estimated VAD & 7.33 & 6.79 & 2.19 \\
\bottomrule
\end{tabular}
\vspace{-5pt}
\end{table}

The results of the different systems evaluated on the official validation set are shown in Table~\ref{tab:exp}. All of our systems substantially outperform the organizer-provided baseline, VibeVoice-ASR\footnote{\url{https://github.com/ASLP-lab/Smart-Glass-Challenge}}. In addition, the SoulX-Transcriber model fine-tuned on the challenge data performs better than the original pretrained version, indicating that a small amount of target-domain data is effective for adapting the model to the acoustic and conversational characteristics of the challenge dataset.

It should be noted that SoulX-Transcriber occasionally suffers from hallucination, which is manifested as repeated token generation or obvious timestamp drift. For System 2 and System 3, audio processed by the frontend modules tends to cause more hallucinations when directly fed into the pretrained SoulX-Transcriber. This may be due to the distortion introduced by frontend processing, as well as the presence of zero-padded non-speech regions, which makes the processed audio distribution different from that of the model's pretraining data. Fine-tuning SoulX-Transcriber and adjusting the decoding configuration can effectively reduce hallucinations. To better reflect the performance of the frontend modules and avoid contaminating the results with hallucinated outputs, for the few remaining problematic samples, we have to abandon full-length audio decoding and instead perform inference on audio segments of approximately 30 seconds. For System 2 and System 3, we evaluate both oracle VAD labels derived from the validation-set annotations and estimated VAD labels produced by our diarization model. Since the timestamps are autoregressively generated by the LLM backend, the DER in experiments using oracle VAD labels may still be non-zero. Compared with SoulX-Transcriber without frontend assistance, System 2 GSS-SoulX and System 3 DPRNN\_mc-SoulX achieve lower error rates. For GSS-SoulX, the oracle-VAD setting performs slightly worse than the estimated-VAD setting, which may be related to the accuracy of the oracle VAD annotations and the output stability of SoulX-Transcriber.

\begin{figure}[t]

\centering
\includegraphics[width=0.9\linewidth]{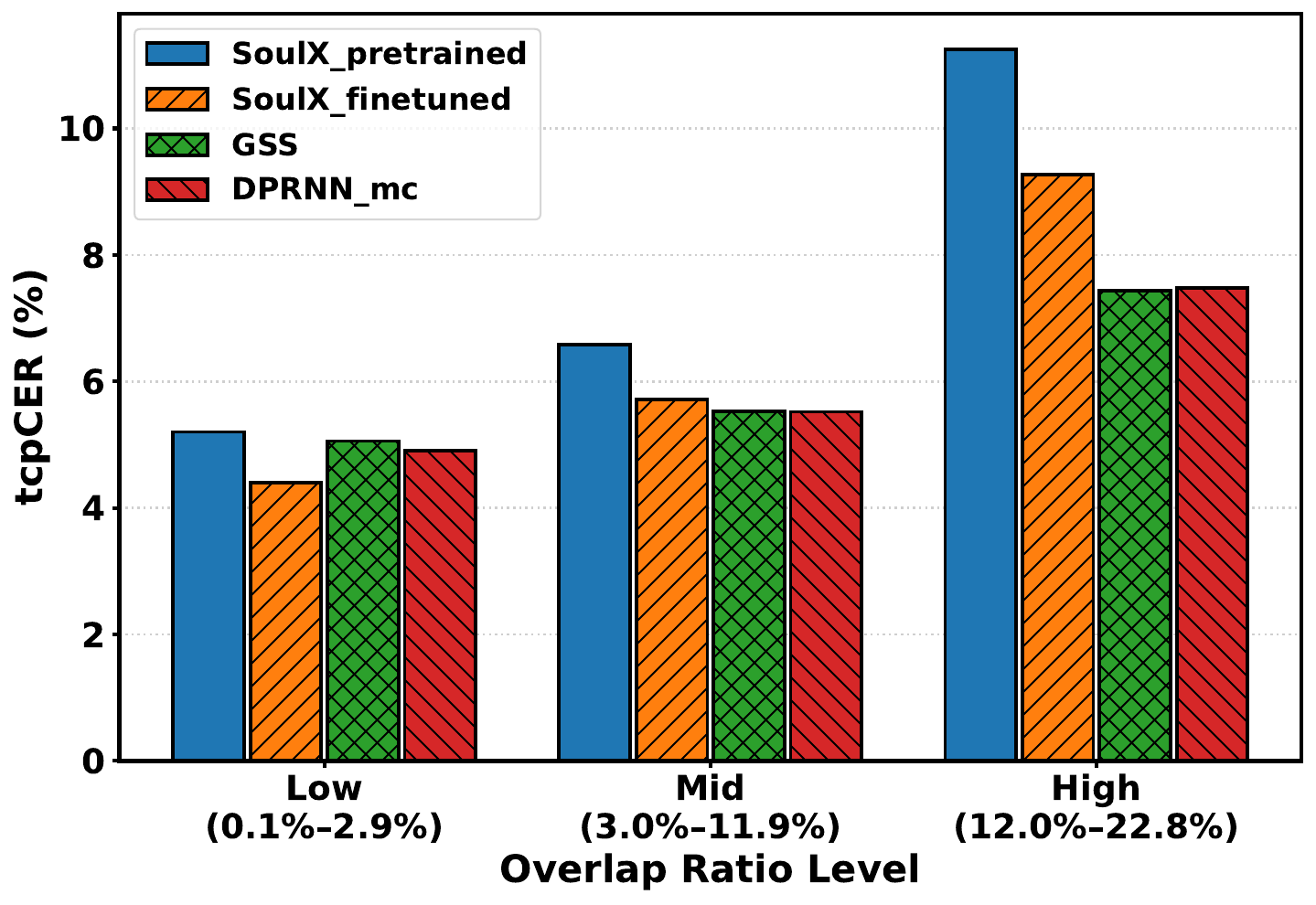}

\caption{Chat-level average tcpCER across different speech overlap ratio levels (Low/Mid/High) for different systems on the Track 1 TSA-ASR task official validation set.}
\label{hist}
\vspace{-5pt}
\end{figure}

Fig.~\ref{hist} shows the chat-level average tcpCER across different speech overlap ratio levels on the validation set for different systems, where System 2 and System 3 use oracle VAD labels. After excluding five chats with annotation errors, we divide the validation set into Low, Mid, and High overlap-ratio groups, containing 26, 25, and 25 chats, respectively, with overlap ratios of 0.1\%--2.9\%, 3.0\%--11.9\%, and 12.0\%--22.8\%. The advantage of speech frontend processing becomes more evident on chats with higher speech overlap ratios. Our final submission is System~1, SoulX-Transcriber-finetuned, as it shows the most stable performance on full-length audio and achieves a tcpCER of 6.22 on the test set.

\section{Conclusion}
The experimental results on the validation set show that frontend-based target-speaker extraction or speech separation, followed by decoding each single-speaker audio track with SoulX-Transcriber, can further reduce CER. This demonstrates the potential of combining speech frontend modules with large audio-language ASR models for speaker-attributed transcription in overlapping speech scenarios.

\clearpage
\bibliographystyle{IEEEtran}
\bibliography{refs}

\end{document}